# Holding spatial solitons in a pumped cavity with the help of nonlinear potentials

Or Maor[1], Nir Dror[2], and Boris A. Malomed[2]

[1] School of Physics and Astronomy, Faculty of Exact Sciences,, Tel Aviv University, Tel Aviv 69978, Israel
[2] Department of Physical Electronics, Faculty of Engineering, Tel Aviv University, Tel Aviv 69978, Israel

We introduce a 1D model of a cavity with the Kerr nonlinearity and saturated gain, designed so as to hold solitons in the state of shuttle motion. The solitons are always unstable in the cavity bounded by the usual potential barriers, due to accumulation of noise generated by the linear gain. Complete stabilization of the shuttling soliton is achieved if the linear barrier potentials are replaced by nonlinear ones, which trap the soliton, being transparent to the radiation. The removal of the noise from the cavity is additionally facilitated by an external ramp potential. The stable dynamical regimes are found numerically, and their basic properties are explained analytically.

The use of spatially localized gain, assisted by local trapping potentials, installed into lossy nonlinear media, has recently drawn considerable attention as a technique for targeted creation and holding of spatial solitons. In the simplest case, the gain is applied as a "hot spot" (HS), on a spatial scale much smaller than the size of the mode to be supported [1-4], hence the HS may be approximated by the $\delta$-function. Settings with multiple HSs [5-9], as well as with extended amplifying structures [10], whose width is comparable to the width of the spatial soliton to be created in them, have been introduced too. These settings can be implemented by implanting gain-producing dopants into one or several narrow segments of the waveguide [11], or by focusing an external pumping beam at the target spot(s) in a uniformly doped waveguide. Solutions for solitons pinned to HSs approximated by $\delta$-functions are available in an analytical form [2,5,9]. More complex modes, such as vortices maintained by the gain applied in an annular area [12-15], have been found numerically.

A natural setting for the implementation of the HS is also offered by lossy multi-core waveguiding arrays, with the gain applied to a single selected core [16]. Remarkably, in the latter case the pinned mode may be stable under the combined action of the *unsaturated* cubic gain and self-defocusing cubic nonlinearity, which is impossible in uniform media (but is possible for localized unsaturated cubic gain acting in a continuous medium [4]).

A problem of obvious significance is stable storage of solitons in amplified cavities [17-22]. The difference from the setting outlined above, where static pinned solitons were considered, is that a narrow soliton should be kept in a broad cavity in the state of stable shuttle oscillations. Our objective here is to propose a one-dimensional (1D) model of a pumped cavity which admits stable storage of the dynamically trapped solitons. Obviously, the simplest possibility is to consider a cavity of size $L$ between $\delta$-functional potential barriers, placed at points $x = \pm L/2$ [23], with saturable gain applied between these points. The corresponding nonlinear Schrödinger (NLS) equation for local amplitude $u(x,z)$ of the electromagnetic field, where $z$ is the propagation distance, is [24,25]

$$iu_z + (1/2)u_{xx} + |u|^2 u = \varepsilon_0 \left[ \delta(x+L/2) + \delta(x-L/2) \right] u$$
$$+ ig_L(x)\left(\Gamma_0 - \Gamma_2 |u|^2\right)u. \qquad (1)$$

Here $\varepsilon_0$ is the height of the barriers, the gain-modulation profile is taken as $g_L(|x|<L/2)=1$, $g_L(|x|>L/2)=0$, while $\Gamma_0$ and $\Gamma_2$ represent the linear gain and nonlinear loss (saturation) acting in the cavity. We fix scales by setting $\Gamma_2 = 0.001$ and $L = 3$ (the exact size of the cavity is not important, provided that it is much larger than the width of the trapped soliton). The $\delta$-functions were approximated by narrow Gaussians with width $\sigma = 0.01$, $\tilde{\delta}(x) = \left(2\pi\sigma^2\right)^{-1/2} \exp\left(-x^2/2\sigma^2\right)$, or, alternatively, by taking $\tilde{\delta}(x) \neq 0$ at a single point of the numerical mesh, the results being the same in both cases. The model may also include linear loss acting outside of the cavity, which is not included in Eq. (1), as it does not essentially affect the results, and absorbers, i.e., stripes with strong linear loss installed at edges of the integration domain. The simulations were performed by means of the split-step fast-Fourier-transform method.

If the cavity is broad enough and parameters $\Gamma_0$ and $\Gamma_2$ are small, one can apply the perturbation theory [26] for the NLS soliton, which is taken as $u(x,z) = \eta \, \text{sech}(x-Cz) \exp\left(iCx + (i/2)(\eta^2 - C^2)z\right)$, with amplitude $\eta$ and tilt (velocity) $C$. Because the linear gain and cubic loss in Eq. (1) do not break the

Galilean invariance, the perturbation theory predicts that $C$ remains a free parameter, while $\eta$ is uniquely selected by the gain-loss balance condition for the total power, $\int_{-\infty}^{+\infty}|u(x,z)|^2\,dx$:

$$\eta_0 = \sqrt{3\Gamma_0/(2\Gamma_2)}. \qquad (2)$$

The analysis of the balance equation for the Hamiltonian, $\int_{-\infty}^{+\infty}\left[|u_x|^2-(1/2)|u|^4\right]dx$ (rather than the total power) produces precisely the same result (2), leaving the tilt free.

However, direct simulations of Eq. (1), supplemented by the linear loss and absorber placed outside of the cavity, demonstrate that the model *never* gives rise to stable shuttle motion of the trapped soliton. The soliton is destabilized via the amplification of random noise by the linear gain in the cavity. The growing noise stays trapped between the potential barriers, and eventually destroys the soliton, see Fig. 1.

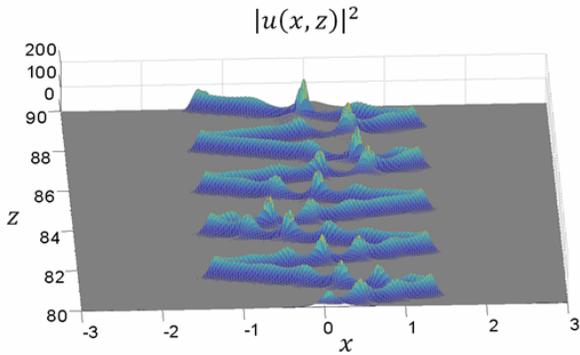

Fig. 1. (Color online) A typical example of the unstable motion of the soliton in the cavity fenced by the $\delta$-functional linear potential barriers with $\varepsilon_0 = 10$, see Eq. (1). The initial tilt (velocity) is $C_0 = 4$, and the gain is $\Gamma_0 = 0.05$. Shown is a stage of the evolution at which the instability is evident.

Systematic simulations also demonstrate that the *sweeping effect*, i.e., periodic suppression of locally growing perturbations by the passing soliton [27], does not stabilize the soliton either. Note that the frequency of the shuttle motion is limited by $f_{\max} = 2L/C_{\max}$, where $C_{\max}$ is the largest tilt at which the moving soliton does not escape the cavity. The comparison of the soliton's kinetic energy, $\eta C^2$ [26], with the height of the effective potential barrier for the soliton, $U_0 = \varepsilon_0 \eta^2$, predicts $C_{\max} = \sqrt{\varepsilon_0 \eta}$, which is corroborated by the simulations.

*Complete stabilization* of the trapped soliton is provided by using *nonlinear potential barriers* instead of the linear ones, represented by terms $\varepsilon_2 \delta(x \mp L/2)|u|^2 u$, instead of $\varepsilon_0 \delta(x \mp L/2)u$, in Eq. (1). Recently, the use of nonlinear potentials in photonics and matter-wave optics has drawn a great deal of attention [28]. In the present context, the nonlinear barriers play the crucial role, as they trap the shuttling soliton, but are *transparent for small-amplitude perturbations*, allowing them to escape the cavity, thus forestalling the growth of the instability through the noise accumulation. The nonlinear potentials may be realized, in particular, by means of narrow doped segments of the waveguide [11,28].

Thus, Eq. (1) is replaced by

$$iu_z + (1/2)u_{xx} + |u|^2 u = \varepsilon_2\left[\delta(x+L/2)+\delta(x-L/2)\right]|u|^2 u$$
$$+ ig_L(x)\left(\Gamma_0 - \Gamma_2|u|^2\right)u + U_{\mathrm{ramp}}(x)u, \qquad (3)$$

which is dealt with hereafter. The last term in Eq. (3) is a linear ramp potential installed outside of the cavity, $U_{\mathrm{ramp}}(|x|>L/2) = -F|x \mp L/2|$, $U_{\mathrm{ramp}}(|x|<L/2) \equiv 0$. It helps to stabilize the trapped soliton by directing the escaping radiation towards the absorbers, rather than allowing it to penetrate back into the cavity. We here set $F = 10$. Simulations demonstrate that very weak instability (much weaker than in Fig. 1) develops in the absence of the ramp, i.e., it is not really necessary unless the propagation length should be very large. The ramp was also included in the simulations of Eq. (1), but did not help to stabilize the soliton in that case.

An example of the completely stable shuttle motion produced by simulations of Eq. (3) is displayed in Fig. 2. After a transient stage, this regime persists indefinitely long [see also Figs. 4 and 5(a) below]. Additional simulations, with random noise added to the input (not shown here in detail), demonstrate that the shuttling soliton remains robust in the latter case as well (the initial noise was introduced, multiplying the input by $1+0.1\cdot\mathrm{Rand}(x)$, with $\mathrm{Rand}(x)$ assuming random values uniformly distributed between $-0.5$ and $+0.5$; in fact, this perturbation is a relatively strong one).

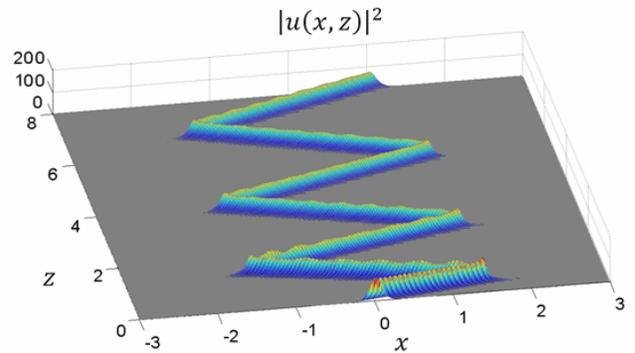

Fig. 2. (Color online) A typical example of the stable shuttle motion of the soliton in the cavity fenced by the *nonlinear* $\delta$-functional barriers with $\varepsilon_2 = 0.1$, see Eq. (3), $C_0$ and $\Gamma_0$ being the same as in Fig. 1.

The stability of the dynamically trapped soliton in Eq. (2) is limited by the possibility of its escape from the cavity if its kinetic energy, $\eta C^2$, exceeds the height of the potential of the repulsion of the soliton by the nonlinear barrier,

$$U(\xi) = (\varepsilon_2 / 2)\eta^4 \text{sech}^4(\eta \xi), \qquad (4)$$

where $\xi$ is the distance of the soliton's center from the edge of the cavity. It follows from here that the soliton escapes at $C > C_{\max} = \sqrt{(\varepsilon_2 / 2)\eta^3}$. Analysis of numerical data corroborates that it is an accurate threshold condition for the passage of the nonlinear barrier, if $\eta$ and $C$, which vary in the course of the evolution, are taken for the soliton hitting the barrier. The numerical data corroborate too that Eq. (2) very accurately predicts the amplitude of the stable soliton, at all values of the gain, as shown in Fig. 3.

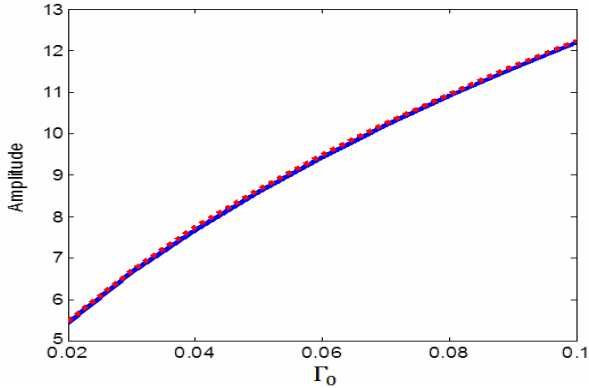

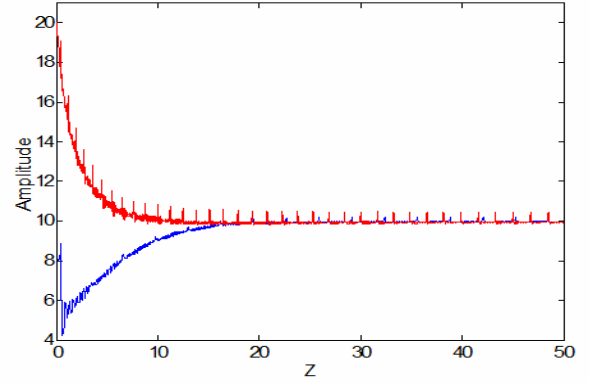

Fig. 4. (Color online) The evolution of soliton's amplitudes toward equilibrium value (2) in simulations of Eq. (3) with $\Gamma_0 = 1/15$. Periodic spikes correspond to transient compression of the soliton hitting the nonlinear potential barriers.

Fig. 3. (Color online) The completely overlapping blue and dashed red curves show amplitudes of the stably shuttling soliton, as found from simulations of Eq. (3), and as predicted by Eq. (2). In agreement with the other prediction of the analysis, the amplitude does not depend on the soliton's tilt (velocity).

The stability of the soliton with amplitude (2) suggests that it is an *attractor* [25] (in terms of the amplitude, while, as concerns the velocity, it is a neutrally stable mode, see below). This is corroborated by Fig. 4, which demonstrates that the amplitude of a soliton, if initially taken larger or smaller than (2), relaxes to this value, except for very small initial amplitudes, for which the soliton is broader than the cavity, which occurs at $\Gamma_0 < 0.02$. On the other hand, if the gain is too large ($\Gamma_0 > 0.1$), the amplified perturbations accumulate too fast, and the transparency of the nonlinear barrier does not provide for complete stabilization of the dynamically trapped soliton. Therefore, dependences of characteristics of the stable dynamical regime on $\Gamma_0$ are displayed, in Figs. 3 and 5(b), in the interval of $0.02 \leq \Gamma_0 \leq 0.1$.

Periodic collisions with the barriers may affect the soliton's velocity (tilt). Indeed, the simulations demonstrate that, if the initial velocity is too large (however, smaller than the above-mentioned escape threshold), it decreases to a smaller value, $C_0(\Gamma_0)$, which is then kept constant, up to the numerical accuracy, see Fig. 5. On the other hand, if the soliton is initially launched with velocity $C < C_0$, it keeps this velocity in the course of the subsequent evolution.

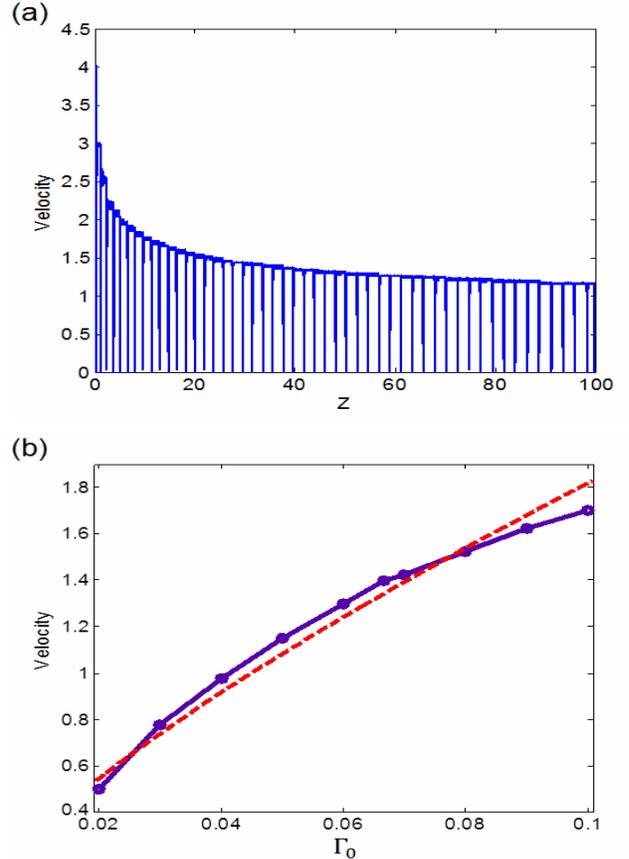

Fig. 5. (Color online) (a) The evolution of the soliton's velocity in the simulations of Eq. (3), with $\Gamma_0 = 0.05$, from an initial value,

$C_{in} = 4$, toward $C_0 = 1.15$. Periodic spikes correspond to the soliton rebounding from the barriers. (b) The upper bound $C_0(\Gamma_0)$ for velocities of stably shuttling solitons with $\varepsilon_2 = 0.1$. The dashed curve is the best fit to the analytically estimated dependence, $C_0 \sim \Gamma_0^{3/4}$.

The dependence of $C_0$ on $\Gamma_0$ can be explained as follows. If the soliton hits the barrier with a relatively high velocity, $C > C_0$, the soliton's tail penetrates into the external area, where it decays into radiation waves driven away by the ramp potential. The ensuing loss of the soliton's power is compensated by the intra-cavity gain, but the gain does not restore the momentum carried away by the radiation waves. This mechanism leads to a small drop of the velocity of the relatively fast soliton, as a result of its rebound from the barrier [as seen in Fig. 5(a)]. Using potential (4) and comparing it to the above-mentioned soliton's kinetic energy, $\eta C^2$, it is possible to estimate the scaling of the dependence of the smallest velocity, at which the soliton can emit radiation outwards, due to the collision with the barrier, on the amplitude: $C_0 \sim \sqrt{\varepsilon_2 \eta^3}$. Finally, in the combination with Eq. (2), this yields $C_0 \sim \Gamma_0^{3/4}$, which is consistent with the numerical findings, as shown in Fig. 5(b).

In conclusion, we have presented the analysis of the model of the pumped cavity for holding spatial solitons in the Kerr medium with cubic losses. In the cavity bounded by the linear $\delta$-functional barriers, the trapped solitons are always unstable, due to the accumulation of noise amplified by the linear gain. The replacement of the barriers by nonlinear ones, which are transparent to the radiation, combined with the external ramp potential, makes the dynamically trapped (shuttling) solitons *completely stable*.

Multi-soliton trapped states are possible too in the present setting, featuring collisions between solitons in the cavity. It may be also interesting to consider trapped *gap solitons* in a pumped cavity based on the spatial Bragg grating. A challenging extension is to develop a 2D version of the present setting.

# References

[1] W. C. K. Mak, B. A. Malomed, and P. L. Chu, "Interaction of a soliton with a localized gain in a fiber Bragg grating", Phys. Rev. E **67**, 026608 (2003).
[2] C.-K. Lam, B. A. Malomed, K. W. Chow, and P. K. A. Wai, "Spatial solitons supported by localized gain in nonlinear optical waveguides", Eur. Phys. J. Special Topics **173**, 233-243 (2009).
[3] Y. V. Kartashov, V. V. Konotop, and V. A. Vysloukh, "Dissipative surface solitons in periodic structures", EPL **91**, 34003 (2010).
[4] O. V. Borovkova, V. E. Lobanov, and B. A. Malomed, "Stable nonlinear amplification of solitons without gain saturation", EPL **97**, 44003 (2012).
[5] C. H. Tsang, B. A. Malomed, C.-K. Lam, and K. W. Chow, "Solitons pinned to hot spots", Eur. Phys. J. D **59**, 81-89 (2010)
[6] D. A. Zezyulin, Y. V. Kartashov, and V. V. Konotop, "Solitons in a medium with linear dissipation and localized gain", Opt. Lett. **36**, 1200-1202 (2011).
[7] Y. V. Kartashov, V. V. Konotop, and V. A. Vysloukh, "Symmetry breaking and multipeaked solitons in inhomogeneous gain landscapes", Phys. Rev. A 83, 041806(R) (2011).
[8] D. A. Zezyulin, V. V. Konotop, and G. L. Alfimov, "Dissipative double-well potential: Nonlinear stationary and pulsating modes", Phys. Rev. E **82**, 056213 (2010).
[9] C. H. Tsang, B. A. Malomed, and K. W. Chow, "Multistable dissipative structures pinned to dual hot spots", Phys. Rev. E **84**, 066609 (2011).
[10] F. K. Abdullaev, V. V. Konotop, M. Salerno, and A. V. Yulin,"Dissipative periodic waves, solitons, and breathers of the nonlinear Schrödinger equation with complex potentials" , Phys. Rev. E **82**, 056606 (2010).
[11] J. Hukriede, D. Runde, and D. Kip, "Fabrication and application of holographic Bragg gratings in lithium niobate channel waveguides," J. Phys. D: Appl. Phys. **36**, R1-R16 (2003).
[12] V. Skarka, N. B. Aleksić, H. Leblond, B. A. Malomed, and D. Mihalache, "Varieties of stable vortical solitons in Ginzburg-Landau media with radially inhomogeneous losses", Phys. Rev. Lett. **105**, 213901 (2010).
[13] V. E. Lobanov, Y. V. Kartashov, V. A. Vysloukh, and L. Torner, "Stable radially symmetric and azimuthally modulated vortex solitons supported by localized gain", Opt. Lett. **36**, 85-87 (2011).
[14] O. V. Borovkova, Y. V. Kartashov, V. E. Lobanov, V. A. Vysloukh, and L. Torner, "Vortex twins and anti-twins supported by multiring gain landscapes", Opt. Lett. **36**, 3783-3785 (2011).
[15] C. Huang, F. Ye, B. A. Malomed, Y. V. Kartashov, and X. Chen, "Solitary vortices supported by localized parametric gain", Opt. Lett. **38**, 2177-2179 (2013).
[16] B. A. Malomed, E. Ding, K. W. Chow and S. K. Lai, "Pinned modes in lossy lattices with local gain and nonlinearity". Phys. Rev. E 86, 036608 (2012).
[17] W. J. Firth, G. K. Harkness, A. Lord, J. M. McSloy, D. Gomila, and P. Colet, "Dynamical properties of two-dimensional Kerr cavity solitons", J. Opt. Soc. Am. B **19**, 747-752 (2002).
[18] P. Mandel and M. Tlidi, "Transverse dynamics in cavity nonlinear optics (2000-2003)", J. Opt. B Quant. Semicl. Opt. **6**, R60-R75 (2004).

[19] A. G. Vladimirov, D. V. Skryabin, G. Kozyreff, P. Mandel, and M. Tlidi, "Bragg localized structures in a passive cavity with transverse modulation of the refractive index and the pump", Opt. Exp. **14**, 1-6 (2006).
[20] Y. Tanguy, T. Ackemann, W. J. Firth, and R. Jaeger, "Realization of a semiconductor-based cavity soliton laser", Phys. Rev. Lett. **100**, 013907 (2008).
[21] M. Tlidi, A. G. Vladimirov, D. Pieroux, and D. Turaev, " Spontaneous motion of cavity solitons induced by a delayed feedback, Phys. Rev. Lett. **103**, 103904 (2009).
[22] C. Yin, D. Mihalache, and Y. He, "Dynamics of two-dimensional dissipative spatial solitons interacting with an umbrella-shaped potential", J. Opt. Soc. Am. B **28**, 342-346 (2011)
[23] P. Y. P. Chen, B. A. Malomed, and P. L. Chu, "Trapping Bragg solitons by a pair of defects", Phys. Rev. E **71**, 066601 (2005).
[24] N. N. Rosanov, *Spatial Hysteresis and Optical Patterns* (Springer, Berlin, 2002).
[25] J. N. Kutz, "Mode-locked soliton lasers", SIAM Rev. **48**, 629–678 (2006).
[26]. Y. S. Kivshar and B. A. Malomed, "Dynamics of solitons in nearly integrable systems", Rev. Mod. Phys. **61**, 763-915 (1989).
[27] M. van Hecke, E. de Wit, and W. van Saarloos, "Coherent and incoherent drifting pulse dynamics in a complex Ginzburg-Landau equation", Phys. Rev. Lett. **75**, 3830-3833 (1995).
[28] Y. V. Kartashov, B. A. Malomed, and L. Torner, "Solitons in nonlinear lattices", Rev. Mod. Phys. **83**, 247-306 (2011).